\documentclass{webofc}

\usepackage[varg]{txfonts}   
\usepackage{hyperref}
\usepackage{url}
\hypersetup{colorlinks=true,citecolor=blue,urlcolor=blue,linkcolor=blue}
\begin{document}
\title{Heavy flavor correlations and Quarkonia production in high energy pp collisions in the EPOS4 framework}
%
%

\author{\firstname{Jiaxing} \lastname{Zhao}\inst{1,2}\fnsep\thanks{Speaker} \and
        \firstname{Taesoo} \lastname{Song}\inst{3} \and
        \firstname{Pol  Bernard} \lastname{Gossiaux}\inst{4}\and
        \firstname{Klaus} \lastname{Werner}\inst{4}\and
        \firstname{Joerg} \lastname{Aichelin}\inst{4}\and
        \firstname{Elena} \lastname{Bratkovskaya}\inst{3,1,2}
}

\institute{Helmholtz Research Academy Hessen for FAIR (HFHF), GSI Helmholtz	Center for Heavy Ion Physics. Campus Frankfurt, 60438 Frankfurt, Germany
\and
           Institute for Theoretical Physics, Johann Wolfgang Goethe University, Max-von-Laue-Str. 1, 60438 Frankfurt am Main, Germany
\and
           GSI Helmholtzzentrum für Schwerionenforschung GmbH, Planckstr. 1, 64291 Darmstadt, Germany  
\and
           SUBATECH, Nantes University, IMT Atlantique, IN2P3/CNRS 4 rue Alfred Kastler, 44307 Nantes cedex 3, France    
          }

\abstract{
In QCD, the production of heavy quark–antiquark pairs can proceed through different mechanisms, each imprinting characteristic correlations between the heavy quarks. In this work, we use the EPOS4 event generator to study how these correlations affect quarkonium production in $pp$ collisions. Our results demonstrate that the observed correlations between heavy mesons directly reflect the underlying production mechanisms and simultaneously shape the transverse-momentum distributions of quarkonia.
}
\maketitle
Quarkonia, bound states of a heavy quark and its antiquark (such as charmonium  and bottomonium), serve as unique and powerful probes of Quantum Chromodynamics (QCD) under extreme conditions and the creation of the quark-gluon plasma (QGP)~\cite{Andronic:2015wma}. Due to the large mass of heavy quarks, their production predominantly occurs in the early stage of high-energy collisions via hard partonic scatterings. Formation of quarkonia related to both the perturbative and non-perturbative aspects of QCD, which has been studied via many approaches, such as the Color-Evaporation Model, Color-Singlet Model, Color-Octet Model, NRQCD and so on.  In recent times, we have advanced a new approach, which is based on the quantum density matrix and which has turned out to give a very good description of the quarkonium production in both $pp$~\cite{Song:2017phm,Zhao:2023dvk} and heavy-ion collisions~\cite{Villar:2022sbv,Song:2023zma}. In this paper, we take charmonium as an example to illustrate the clear connection between quarkonium production and the correlations of heavy quark–antiquark pairs. This study is carried out using the EPOS4 event generator (version EPOS4.0.1.s9), which incorporates different $Q\bar Q$ production processes.

Based on the quantum density formalism, the probability that a meson $i$ is produced is given by $P_i= Tr (\rho_i \rho^{(N)})$ with $\rho_i$ being the density matrix of the meson $i$ and $\rho^{(N)}$ the density matrix of the $N$ heavy quarks and antiquarks, such as produced in a $pp$ collision. A partial Fourier transformation of the density matrices yields, 
\begin{eqnarray}
\frac{dP_i}{{d^3\bf R}{d^3\bf P}}=\sum \int {d^3rd^3p \over (2\pi)^6}W_{i}^\Phi({\bf r},{\bf p})\prod_{j>2} \int {d^3r_jd^3p_j \over (2\pi)^{3(N-2)}}W^{(N)}({\bf r}_1,{\bf p}_1,...,{\bf r}_N,{\bf p}_N),
\label{eq.projection}
\end{eqnarray}
where $W_i^\Phi$ is the quarkonium Wigner density and $W^{(N)}({\bf r}_1,{\bf p}_1,...,{\bf r}_N,{\bf p}_N)$ is the quantal density matrix in Wigner representation of the ensemble of $N$ heavy quarks  produced in a $pp$ collision. ${\bf r}$ $({\bf R})$ and ${\bf p}$ $({\bf P})$ are the relative (center of mass) coordinate and momentum of the heavy quark and antiquark pairs. We assume that the unknown quantal $N$-body Wigner density can be replaced by the average of classical phase space distributions, $W^{(N)}\approx \langle W^{(N)}_{\rm classical} \rangle$. The classical momentum space distributions of the heavy quarks is provided by EPOS4~\cite{Werner:2023zvo,Werner:2023fne}.  However, only the coordinate information of the vertex where the $Q\bar Q$ pair is created is provided. For a heavy quark pair, created at the same vertex, we assume that the relative distance between $Q$ and $\bar Q$  in their center-of-mass frame is given by a Gaussian distribution, $r_{cm}^2\exp (- r_{cm}^2 /( 2\sigma_{\rm Q\bar Q})^2 )$, where the relative distance is controlled by the effective width $\sigma_{\rm Q\bar Q}$, which is chosen, $\sigma_{c\bar c}=0.35~\rm fm$, for all charmonium states.
\begin{figure}[h]
\centering
\includegraphics[width=12cm,clip]{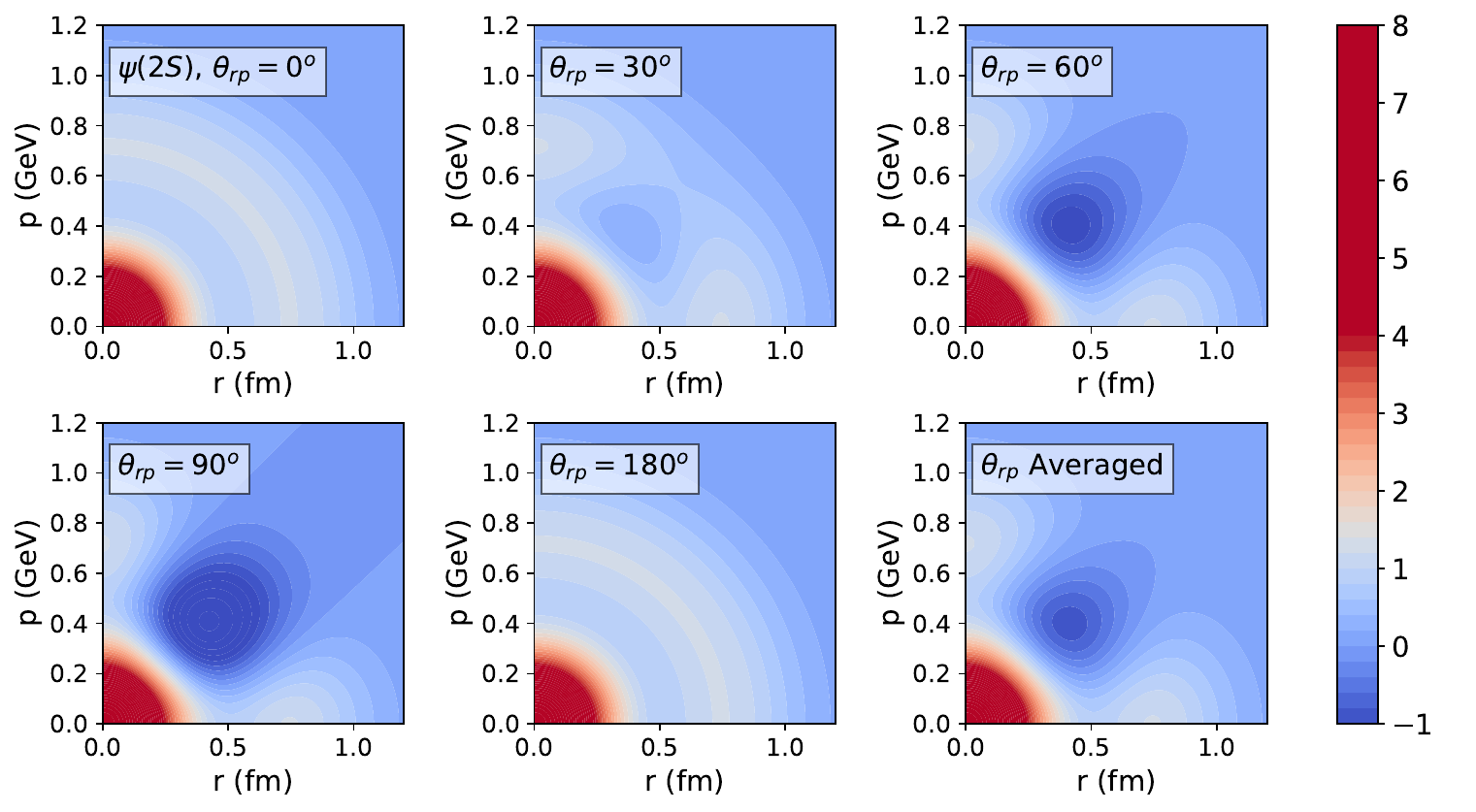}
\caption{Wigner density of $\psi(2S)$. Different panel shows the angle $\theta_{rp}$-dependent and averaged Wigner density. The width $\sigma=0.452$ fm for $\psi(2S)$~\cite{Zhao:2023dvk}.}
\label{fig-1}
\end{figure}

\begin{figure}[h]
\centering
\includegraphics[width=12cm,clip]{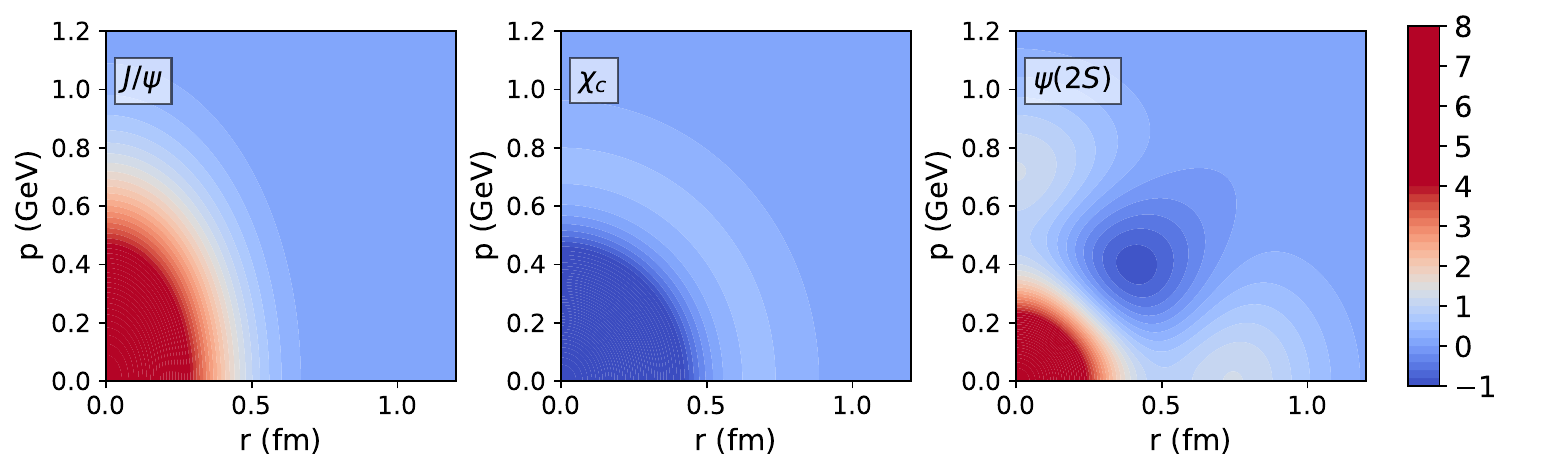}
\caption{Wigner densities of different charmonium  states. The width $\sigma=$ 0.348, 0.426, and 0.452 fm for $J/\psi$, $\chi_c$, and $\psi(2S)$, respectively~\cite{Zhao:2023dvk}.}
\label{fig-2}
\end{figure}

The vacuum properties of quarkonia can be well described by the relativistic or non-relativistic potential model~\cite{Zhao:2020jqu}. 
Aim to get a analytical Wigner density, one usually assume that the potential between the heavy quark pairs can be approximated by a 3-D isotropic harmonic oscillator potential. Then, the Wigner density can be constructed via a Wigner transformation in spherical coordinates. For any given quarkonium state $\Phi$, the Wigner density can be found in Ref.~\cite{Zhao:2023dvk}. For charmonia,
\begin{eqnarray}
W_{\rm J/\psi}({\bf r,p})&=&8e^{-\xi} ,\nonumber\\
W_{\rm \chi_c}({\bf r,p})&=& {8\over 3}e^{-\xi }\Big(2\xi -3 \Big),\nonumber\\
W_{\rm \psi(2S)}({\bf r,p})&=& {8\over 3}e^{-\xi}\Big(3+2\xi^2-4\xi-8[p^2r^2-({\bf p}\cdot{\bf r})^2)] \Big),
\end{eqnarray}
where $\xi=r^2/\sigma^2+p^2 \sigma^2$.
The Wigner density of $\psi(2S)$ depends not only on $r=|{\bf r}|$ and $p=|\bf p|$, but also on the angle $\theta_{rp}$ between $\bf r$ and $\bf p$, as clearly shown in Fig.~\ref{fig-1}. If averaged over the angle $\theta_{rp}$, the Wigner densities can be shown as a function of $r$ and $p$ in Fig.~\ref{fig-2}. 
The width in the Wigner density is related to the root-mean-radius of the quarkonium, as shown in Ref.~\cite{Zhao:2023dvk}. Similar results are also shown in Ref.~\cite{Kordell:2021prk}.

\begin{figure}[h]
\centering 
\includegraphics[width=4.5cm,clip]{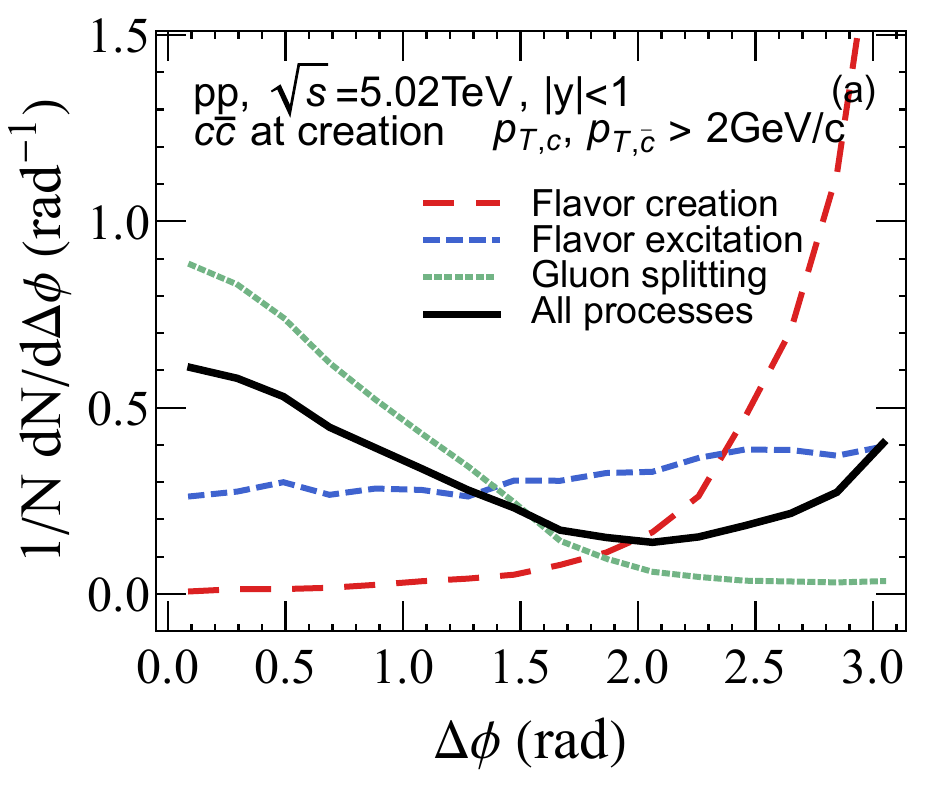} \includegraphics[width=4.5cm,clip]{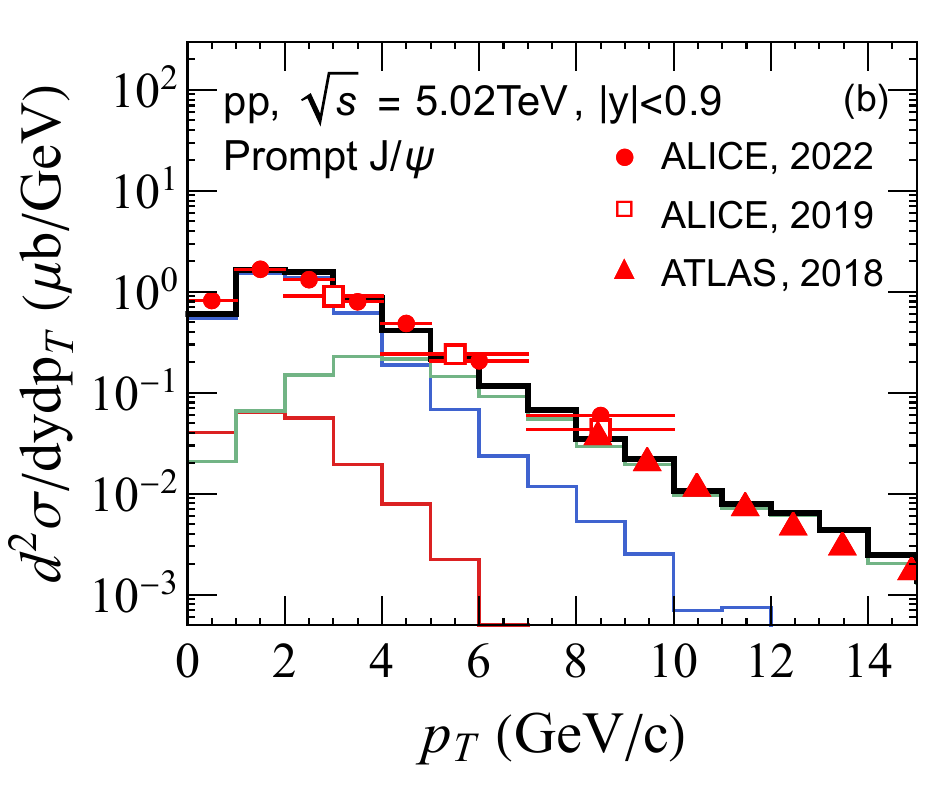}\\
\includegraphics[width=4.5cm,clip]{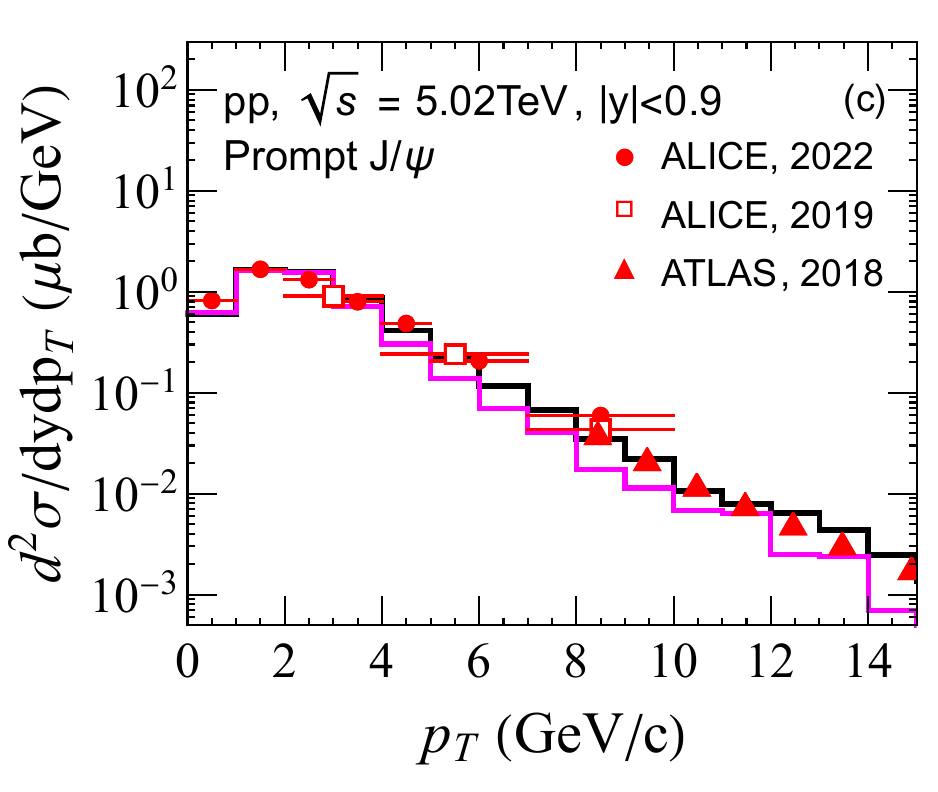} \includegraphics[width=4.5cm,clip]{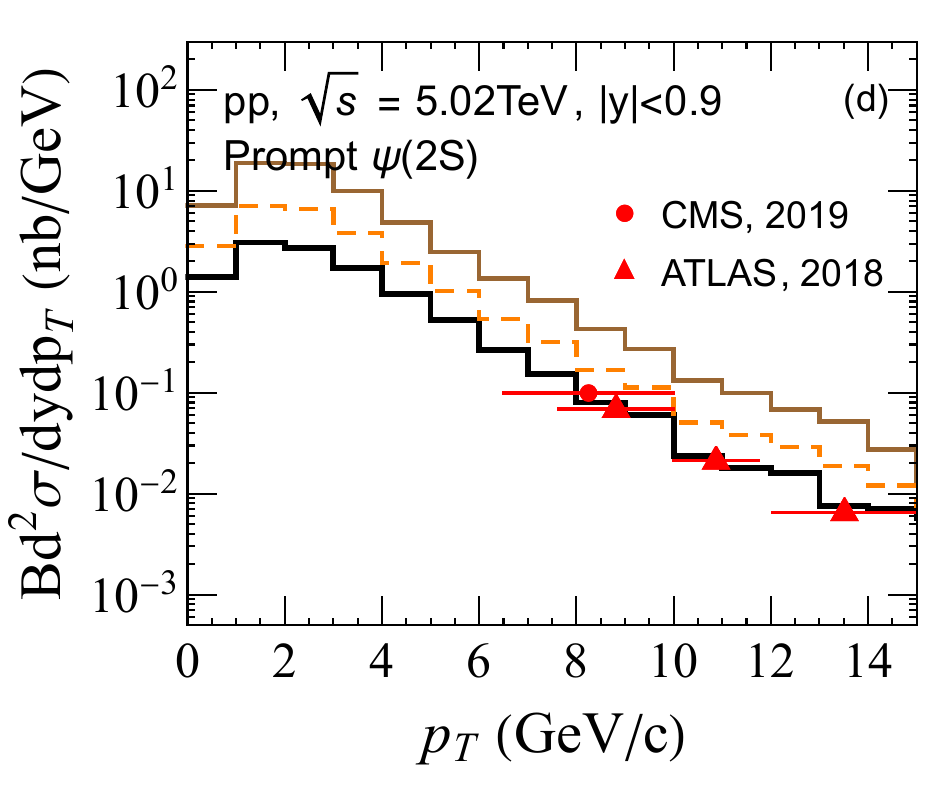}
\caption{ The correlation between $c$ and $\bar c$ (a). $p_T$ spectra of prompt $J/\psi$ (b) in $pp$ collisions at $\sqrt{s}=5.02~\mathrm{TeV}$, where red, blue, and green curves correspond to flavor creation, flavor excitation, and gluon splitting, respectively. The black solid line denotes their total contributions. The magenta line in (c) represents the result obtained without correlations between $c$ and $\bar c$. (d) shows the $p_T$ spectra of $\psi(2S)$ with the $\theta_{rp}$-dependent Wigner density.}
\label{fig-3}
\end{figure}

In EPOS4, $Q\bar Q$ pairs can be produced through three leading-order or next-to-leading-order processes: flavor creation, flavor excitation, and gluon splitting~\cite{Werner:2023fne}. Flavor creation leads to a back-to-back correlation, but its overall contribution is relatively small. Gluon splitting dominates charm production at high $p_T$ and results in a same-direction correlation with a small azimuthal opening angle. Flavor excitation, typically associated with gluon emission carrying significant momentum, produces almost no visible correlation between $c$ and $\bar c$, as illustrated in Fig.\ref{fig-3}-a. The $c\bar c$ correlation can be measured directly in the experiment, for example, via $D\bar D$ correlations~\cite{LHCb:2012aiv}.
Charmonia are formed via the Wigner projection method, and the corresponding results are presented in Fig.~\ref{fig-3}-b. We observe that $J/\psi$ with $p_T \gtrsim 4$ GeV are produced almost exclusively from charm quarks originating in gluon-splitting processes (green line), where the $c$ and $\bar c$ quarks remain close in phase-space, creating favorable conditions for charmonium formation. If such correlations are ignored (preserve the $p_T$ of each particle but randomize the relative angle between them), charmonium production at high $p_T$ is clearly suppressed, leading to an underestimation relative to experimental data as shown in Fig.\ref{fig-3}-c. 
Finally, we examine the influence of the angle-dependent Wigner density for $\psi(2S)$. In the calculation, we consider $\theta_{rp}=0^\circ$ (brown), $60^\circ$ (orange-dashed), $90^\circ$ (fully negative, therefore not shown), and the $\theta_{rp}$-averaged (black) Wigner density. The results (Fig.~\ref{fig-3}-d) show a clear difference. While such an angle-dependent Wigner density can be neglected in $pp$ collisions, it may become important for excited-state production in anisotropic systems.


In conclusion, our results show that the measured heavy-meson correlations provide a direct fingerprint of the underlying production mechanisms, each leaving a distinct imprint on the transverse-momentum distributions of quarkonia in high energy $pp$ collisions.

\vspace*{1mm}
\textit{Acknowledgements:}
We acknowledge support by the Deutsche Forschungsgemeinschaft (DFG, German Research Foundation) through the grant CRC-TR 211 'Strong-interaction matter under extreme conditions'-Project number 315477589-TRR 211. 
%
\bibliography{referencesqm} 
%
%
%
%

\end{document}